\documentclass[12pt]{article}
\usepackage[centertags]{amsmath}
\usepackage{amsfonts}
\usepackage{amssymb}
\usepackage{amsthm} 
\usepackage{newlfont}
\usepackage{epsfig}
\usepackage{amscd}

\newcommand{\beq}{\begin{equation}}
\newcommand{\eeq}{\end{equation}}
\newcommand{\ba}{\begin{array}}
\newcommand{\ea}{\end{array}}
\newcommand{\bea}{\begin{eqnarray}}
\newcommand{\eea}{\end{eqnarray}}

\begin{document}

\begin{center}
{\large \sc \bf A modification of the method of characteristics: a new class of multidimensional partially integrable nonlinear systems}

\vskip 15pt

{\large A. I. Zenchuk }

\vskip 8pt

\smallskip

{\it $^2$ Institute of Chemical Physics, RAS
Acad. Semenov av., 1
Chernogolovka,
Moscow region
142432,
Russia}

\smallskip

\vskip 5pt


\vskip 5pt

{\today}

\end{center}

\begin{abstract}
We represent an algorithm  reducing a big class of  systems of ($M+1$)-dimensional nonlinear   partial differential equations (PDEs)  to the systems of $M$-dimensional first order PDEs. Thus, we integrate the original system with respect to only one independent variable reducing its dimensionality by one. For this reason we call such systems partially integrable ones. 
In particular, if $M=1$, then $M$-dimensional PDEs become ODEs. 
Our method may be referred to as a generalization of the method of characteristics. Possible application in hydrodynamics is discussed.
\end{abstract}

\section{Introduction}
\label{Section:first_order}
The method of characteristics is known as an effective method for solving the first order scalar multidimensional  PDEs reducing them  to the appropriate  systems of  scalar  ODEs \cite{KG}.
Regarding the systems of scalar multidimensional first order PDEs, their integration is more sophisticated.  There are  few  known classes of integrable systems among them \cite{KG,T,DN,T2,F,K}. We underline only one class of such systems which is most relevant to the algorithm developed in our letter, namely,  the systems of ($M+1$)-dimensional  PDEs with the same leading parts \cite{KG}:
\begin{eqnarray}\label{cl_lead}
u^{(j)}_t+\sum_{i=1}^M a_i u^{(j)}_{x_i} - b_j =0,\;\;j=1,\dots,N,
\end{eqnarray}
 where $a_i$ and $b_j$ are arbitrary functions of $t$, $x_i$, $i=1,\dots,M$ 
and $u^{(j)}$, $j=1,\dots,N$.   Thus, $t$ and $x_i$, $i=1,\dots,M$ are independent variables while $u^{(i)}$, $i=1,\dots,N$ are scalar fields.
 It is  shown (see \cite{KG}) that  the solutions $u^{(i)}$  to the system of nonlinear PDEs  (\ref{cl_lead})
may be found as solutions to the system of non-differential equations
\begin{eqnarray}\label{FF}
F^{(k)}(t,x,u) =0,\;\;k=1,\dots,N,
\end{eqnarray}
where  $F^{(k)}$ are solutions to the following linear PDE:
\begin{eqnarray}\label{F_lin_cl}
F^{(k)}_t + \sum_{i=1}^M a_i F^{(k)}_{x_i} +\sum_{i=1}^N b_i F^{(k)}_{u^{(i)}} =0,\;\;\;\left.\frac{\partial(F^{(1)}\dots F^{(N)})}{\partial(u^{(1)},\dots,u^{(n)})}\right|_{{\mbox{Eqs.(\ref{FF})}}}\neq 0.
\end{eqnarray}
We show that the algorithm based on  equations (\ref{FF}) and (\ref{F_lin_cl}) may be generalized introducing a   new type of partially integrable systems of  multidimensional nonlinear PDEs having the following general form:
\begin{eqnarray}\label{intr_gen}
L \vec v + \sum_{m=1}^M \hat A^{(m)} L \vec v_{x_m}  = \vec A,
\end{eqnarray}
where $L=\partial_t + \sum_{m=1}^M h^{(m)} \partial_{x_m}$, 
$\vec v=(v^{(1)},\dots,v^{(N)})^T$ is a vector field, $\hat A^{(m)}$, $\vec A$ 
and $h^{(m)}$ ($m=1,\dots,M$)
are $N\times N$ matrices, $N$-dimensional vector and scalars respectively  depending on $v^{(i)}$ and $v^{(i)}_{x_j}$, $i=1,\dots,N$, $j=1,\dots,M$. Superscript  $T$ means matrix transpose.
 Here "partially integrable systems" means that we are able to reduce the system of ($M+1$)-dimensional scalar nonlinear PDEs to the appropriate  system of $M$-dimensional PDEs, i.e. one can  reduce the dimensionality of the original nonlinear PDEs by one. In particular, if $M=1$, then we reduce the two-dimensional system of PDEs to the system of  ODEs. Emphasize, that, in general, our algorithm provides $N$ arbitrary functions of $M$ independent variables in the solution space of considered nonlinear PDEs, i.e. we deal with the general solution. This conclusion is justified in the  examples of  Sec.\ref{Section:examples}, where the initial value problem for the derived nonlinear systems is investigated.

Among the equations of new type we point on the following (2+1)-dimensional system of two PDEs for scalar  fields $u$ and $w$:
\begin{eqnarray}\label{intr_nl1}
&&
u_t-\nu(u_{x_1}+i u_{x_2}) +\frac{1}{2}(u_{x_1}+u_{x_2})^2 +w^3+ (u_{x_1}-i u_{x_2})_t +\\\nonumber
&& +
w \Big(
\nu(u_{x_1}+i u_{x_2})_t-(u_{x_1}+u_{x_2})(u_{x_1}+ u_{x_2})_t
\Big)=0,\\\nonumber
&&
w_t+1=0,\;\;\;\nu=const,
\end{eqnarray}
We show that the construction of the general solution to this (2+1)-dimensional system is reduced to the solution of the system of two  first order two-dimensional nonlinear  PDEs:
\begin{eqnarray}\label{intr_H}
H^{(k)}(w+t,u,u_{x_1},u_{x_2},t)+C^{(k)}(x_1,x_2)=0,
\end{eqnarray}
where $C^{(k)}$ are arbitrary functions of two variables and the functions $H^{(k)}$ satisfy the following system of linear PDEs:
\begin{eqnarray}\label{intr_Hd}
H^{(k)}_t(z,u,p^{(1)},p^{(2)},t) &=& \\\nonumber
&&\hspace{-2cm}
H^{(k)}_u(z,u,p^{(1)},p^{(2)},t) 
\Big(-\nu (p^{(1)} + i p^{(2)}) + \frac{1}{2} (p^{(1)}+p^{(2)})^2 + (z-t)^3\Big),\\\nonumber
H^{(k)}_{p^{(1)}}(z,u,p^{(1)},p^{(2)},t)&=& H^{(k)}_u(z,u,p^{(1)},p^{(2)},t) \Big((\nu-p^{(1)}-p^{(2)})(z-t) +1\Big),\\\nonumber
H^{(k)}_{p^{(2)}}(z,u,p^{(1)},p^{(2)},t)&=& H^{(k)}_u(z,u,p^{(1)},p^{(2)},t) \Big((i\nu-p^{(1)}-p^{(2)})(z-t) -i\Big),\\\nonumber
\end{eqnarray}
 where $k=1,2$. Nonlinear PDEs (\ref{intr_H}) ($a$) have derivatives with respect to the independent variables $x_1$ and $x_2$ while variable $t$ appears as a parameter and ($b$) involve two arbitrary functions of variables $x_1$ and $x_2$.
As a consequence, the initial value problem may be  correctly formulated for system (\ref{intr_nl1}). The peculiarity of system (\ref{intr_nl1}) is that the properly introduced  small scale expansion reduces this system to the following single PDE:
\begin{eqnarray}\label{intr}
v_\tau + v(v_{y_1}+v_{y_2}) - \nu (v_{y_1y_1}+v_{y_2y_2})=0,
\end{eqnarray}
 see Sec.\ref{Section:example_hydr}. Eq.(\ref{intr})  describes, for instance, 
the special case of the dynamics of two-dimensional viscous incompressible fluid with the constant kinematic viscosity $\nu$ when two velocity components are equal to each other.
Eq.(\ref{intr}) becomes (1+1)-dimensional nonlinear B\"urgers equation (integrable by the Hopf-Cole substitution \cite{BLR,H,C}) if either $u_{x}=0$ or $u_{y}=0$. 

The general algorithm for deriving  the new type of systems of  nonlinear PDEs is given in Sec.\ref{Section:Gen}.
Several examples of such systems are represented in Sec.\ref{Section:examples}. Concluding remarks are  given in Sec.\ref{Section:conclusions}.

\section{General algorithm}
\label{Section:Gen}
\subsection{Derivation of partially integrable systems}
\label{Section:gen}
Let us consider the $(M+1)$-dimensional space of variables $t$ and $x_i$, $i=1,\dots,M$,  which are independent variables of the nonlinear PDEs written for the functions $u^{(i)}$, $i=1,\dots,M$ and $w^{(i)}$, $i=1,\dots,K$.  Hereafter we will use the notations $p^{(ij)}=u^{(i)}_{x_j}$,
\begin{eqnarray}\label{note}
&&
u=\{u^{(i)}:\;i=1,\dots,N\},\;\;w=\{w^{(l)}:\;l=1,\dots,K\},\;\;
x=\{x_j: \;j=1,\dots,M\},\\\nonumber
&&
p=\{p^{(ij)}\equiv u^{(i)}_{x_j}:\; i=1,\dots,N,\;\;j=1,\dots,M\}.
\end{eqnarray}

{\bf Theorem 1.}
Let the functions  $F^{(k)}(t,x,u,w,p)$ ($k=1,\dots, M+K$)  of  ($M+N+K+NM +1$) variables  satisfy the following overdetermined  system of linear PDEs:
\begin{eqnarray}\label{Ft}
E^{(k)}&:=&F^{(k)}_t +\sum_{j=1}^M F^{(k)}_{x_j} h^{(j)} = \sum_{i=1}^N F^{(k)}_{u^{(i)}} A^{(i)} + 
 \sum_{l=1}^K F^{(k)}_{w^{(l)}} \tilde A^{(l)} ,\\\label{Fz}
E^{(nmk)}&:=&
F^{(k)}_{p^{(nm)}}= \sum_{i=1}^N F^{(k)}_{u^{(i)}} A^{(nmi)} + 
\sum_{l=1}^K F^{(k)}_{w^{(l)}} \tilde A^{(nml)} ,
\end{eqnarray}
which must be compatible. Here $h^{(j)}$, $A^{(i)}$, $\tilde A^{(l)}$, $A^{(nmi)}$, $\tilde A^{(nml)}$, $\forall \; i,j,l,n,m$, are functions of $t$, $x$, $u$, $w$ and $p$.
Let the functions $u^{(i)}$ ($i=1,\dots, N$) and $w^{(l)}$ ($l=1,\dots, K$) be solutions 
to the following system of equations:
\begin{eqnarray}\label{F}
&&
F^{(k)}(t,x,u,p,w)=0,\;\;k=1,\dots,N+K,
\end{eqnarray}
where
\begin{eqnarray}\label{wr}
\left.\frac{\partial  (F^{(1)}\dots F^{(N+K)})}{\partial (u^{(1)}\dots u^{(N)}w^{(1)}\dots w^{(K)})}\right|_{{\mbox{Eqs.(\ref{F})}}} \neq 0.
\end{eqnarray}
Then the functions $u$ and $w$ are solutions to the following system of nonlinear ($M+1$)-dimensional PDEs:
\begin{eqnarray}\label{nleq1}
&&
u^{(i)}_t+\sum_{n=1}^M u^{(i)}_{x_n}h^{(n)} + A^{(i)}+\sum_{j=1}^N\sum_{m=1}^M A^{(jmi)} 
\left(u^{(j)}_{x_m t} + \sum_{n=1}^M u^{(j)}_{x_m x_n} h^{(n)}\right)=0,\\\label{nleq2}
&&
w^{(l)}_{t}+\sum_{n=1}^M w^{(l)}_{x_n} h^{(n)}+ \tilde A^{(l)}+\sum_{j=1}^N\sum_{m=1}^M \tilde A^{(jml)} 
\left(u^{(j)}_{x_m t} + \sum_{n=1}^M u^{(j)}_{x_m x_n} h^{(n)}\right)=0
\end{eqnarray}
(which is analogous to  eq.(\ref{intr_gen})).

{\it Proof.}
First of all, let us 
differentiate eq.(\ref{F})  with respect to $t$ and $x_n$, $n=1,\dots,M$:
\begin{eqnarray}\label{DFt}
{\cal{E}}^{(k)}&:=&\sum_{i=1}^N F^{(k)}_{u^{(i)}} u^{(i)}_t + \sum_{i=1}^N\sum_{j=1}^MF^{(k)}_{p^{(ij)}} p^{(ij)}_t + \sum_{l=1}^K F^{(k)}_{w^{(l)}} w^{(l)}_t+ F^{(k)}_t =0,\\\label{DFx}
{\cal{E}}^{(nk)}&:=&\sum_{i=1}^N F^{(k)}_{u^{(i)}} u^{(i)}_{x_n} + \sum_{i=1}^N\sum_{j=1}^MF^{(k)}_{p^{(ij)}} p^{(ij)}_{x_n}+ \sum_{l=1}^K F^{(k)}_{w^{(l)}} w^{(l)}_{x_n} + F^{(k)}_{x_n} =0.
\end{eqnarray}
Consider the following combination of equations (\ref{DFt}) and (\ref{DFx}):
\begin{eqnarray}\label{der_nl10}
&&
{\cal{E}}^{(k)}+\sum_{j=1}^M {\cal{E}}^{(jk)} h^{(j)}=0,\;\;k=1,\dots,N+K.
\end{eqnarray}
Eliminating derivatives $F^{(k)}_t$ and $F^{(k)}_{p^{(nm)}}$ from system (\ref{der_nl10}) using eqs.(\ref{Ft}) and (\ref{Fz})
one obtains the following linear combination of the derivatives $F^{(k)}_{u^{(i)}}$ and 
$F^{(k)}_{w^{(i)}}$:
\begin{eqnarray}\label{der_nl1}
&&
\sum_{i=1}^N F^{(k)}_{u^{(i)}}\left (u^{(i)}_t+\sum_{n=1}^M u^{(i)}_{x_n}h^{(n)} + A^{(i)}+\sum_{j=1}^N\sum_{m=1}^M A^{(jmi)} 
\left(p^{(jm)}_t + \sum_{n=1}^M p^{(jm)}_{x_n} h^{(n)}\right)\right) +\\\nonumber
&&
 \sum_{i=1}^K F^{(k)}_{w^{(i)}} \left (w^{(i)}_{t}+\sum_{n=1}^M w^{(i)}_{x_n} h^{(n)}+ \tilde A^{(i)}+\sum_{j=1}^N\sum_{m=1}^M \tilde A^{(jmi)} 
\left(p^{(jm)}_t + \sum_{n=1}^M p^{(jm)}_{x_n} h^{(n)}\right)\right)=0,
\end{eqnarray}
In view of condition (\ref{wr}),
 all coefficients ahead of $F_{u^{(i)}}$ and $F_{w^{(i)}}$ in eq.(\ref{der_nl1})  must be zero which yields
 the  system of nonlinear PDEs (\ref{nleq1},\ref{nleq2}). $\Box$

{\bf Remark:}
It is obvious, that the system (\ref{F}) is a system of  $M$-dimensional PDEs, in general, due to the functions $p^{(ij)}=u^{(i)}_{x_j}$, $j=1,\dots, M$. Variable $t$ appears as a parameter in eqs.(\ref{F}). Thus we reduce  the dimensionality of  original $(M+1)$-dimensional nonlinear PDEs  (\ref{nleq1},\ref{nleq2}) by one. It is important that  the solution space to nonlinear PDEs does admit the ($N+K$) arbitrary functions of $M$ variables in general case.
In fact, eqs.(\ref{Ft},\ref{Fz}) mean that functions $F^{(k)}$ are arbitrary functions of variables $x_i$,
 $u^{(j)}$ and  $w^{(k)}$ ($i=1,\dots,M$, $j=1,\dots,N$, $k=1,\dots,K$). In view of condition (\ref{wr}), one can resolve eqs.(\ref{F}) with respect to $u^{(j)}$ and $w^{(k)}$  and
rewrite the system (\ref{F}) in the following form:
\begin{eqnarray}\label{arb_f}
u^{(j)}&=&\tilde F^{(1,j)}(t,x,p),\;\;\;j=1,\dots,N,\\\nonumber
w^{(k)}&=&\tilde F^{(2,k)}(t,x,p),\;\;\;k=1,\dots,K,
\end{eqnarray}  
where ($N+K$) functions $\tilde F^{(i,j)}$ keep arbitrary dependence on $M$  variables $x^{(i)}$, $i=1,\dots,M$.
The availability of arbitrary functions  allows one to study, for instance, the initial value problem for  nonlinear PDEs (\ref{nleq1},\ref{nleq2}). This is reflected in examples of Sec.\ref{Section:examples}.

We see that the structure of the system of nonlinear PDEs  (\ref{nleq1},\ref{nleq2}) is defined by the coefficients in eqs.(\ref{Ft},\ref{Fz}), i.e. by the functions $A^{(i)}$, $A^{(lmi)}$, $\tilde A^{(j)}$, $\tilde A^{(lmj)}$ and $h^{(j)}$. These functions may not be arbitrary since they  must provide the compatibility of the overdetermined system (\ref{Ft}) and (\ref{Fz}). This problem is discussed in the next subsection.

\subsection{The compatibility conditions of eqs.  (\ref{Ft}) and (\ref{Fz})}
\label{Section:comp}

The compatibility conditions of eqs.  (\ref{Ft}) and (\ref{Fz})  read: $E^{(nmk)}_t=E^{(k)}_{p^{(nm)}}$, $n=1,\dots,N$, $m=1,\dots,M$, $k=1,\dots,N+K$, or
\begin{eqnarray}\label{comp1}
\sum_{i=1}^N F^{(k)}_{u^{(i)}} Q^{(nmi)} + \sum_{l=1}^K F^{(k)}_{w^{(l)}} \tilde Q^{(nml)} + \sum_{j=1}^M F^{(k)}_{x_j} q^{(nmj)}=0,
\end{eqnarray}
where
\begin{eqnarray}\label{c11}
Q^{(nmi)}&=&A^{(i)}_{p^{(nm)}} - A^{(nmi)}_t -
\sum_{j=1}^M A^{(nmi)}_{x_j} h^{(j)} +
\sum_{\tilde i=1}^N \left(
A^{(nmi)}_{u^{(\tilde i)}} A^{(\tilde i)} -A^{(i)}_{u^{(\tilde i)}} A^{(nm\tilde i)} 
\right)+\\\nonumber
&&
\sum_{\tilde l=1}^K \left(
A^{(nmi)}_{w^{(\tilde l)}} \tilde A^{(\tilde l)} -A^{(i)}_{w^{(\tilde l)}} \tilde A^{(nm\tilde l)} 
\right),
\\\label{tc11}
\tilde Q^{(nml)}&=&\tilde A^{(l)}_{p^{(nm)}} - \tilde A^{(nml)}_t -
\sum_{j=1}^M \tilde A^{(nml)}_{x_j} h^{(j)} +
\sum_{\tilde i=1}^N \left(
\tilde A^{(nml)}_{u^{(\tilde i)}} A^{(\tilde i)} -\tilde A^{(l)}_{u^{(\tilde i)}} A^{(nm\tilde i)} 
\right)+\\\nonumber
&&
\sum_{\tilde l=1}^K \left(
\tilde A^{(nml)}_{w^{(\tilde l)}} \tilde A^{(\tilde l)} -\tilde A^{(l)}_{w^{(\tilde l)}} \tilde A^{(nm\tilde l)} 
\right),
\\
\label{hc11}
q^{(nmj)}&=&
h^{(j)}_{p^{(nm)}}  -
\sum_{\tilde i=1}^N h^{(j)}_{u^{(\tilde i)}} A^{(nm\tilde i)} 
-
\sum_{\tilde l=1}^K h^{(j)}_{w^{(\tilde l)}} \tilde A^{(nm\tilde l)} 
.
\end{eqnarray}
If at least some coefficients are nonzero in eq.(\ref{comp1}), then this equation exhibits a linear dependence among the derivatives 
$F_{u^{(i)}}$, $ F_{w^{(l)}} $ and $ F_{x_j}$, $\forall$ $i,j,l$. Since  derivatives $F_{u^{(i)}}$ and $ F_{w^{(l)}} $ are linearly independent (in view of condition (\ref{wr})), 
the above linear dependence would reduce  the dimensionality of nonlinear PDEs, which is not 
desirable. 
 Thus it is reasonable to require that all coefficients in eq.(\ref{comp1}) are zeros:
\begin{eqnarray}\label{eqs_for_A1}
&&
Q^{(nmi)}= \tilde Q^{(nml)}=0,\\\label{eqs_for_A12}
&&
q^{(nmj)}=0, \;\;i,n=1,\dots,N,\;\;j,m=1,\dots,M,\;\;l=1,\dots,K.
\end{eqnarray}
Equations (\ref{eqs_for_A1},\ref{eqs_for_A12}) represent the first system of nonlinear PDEs defining the functions $A^{(i)}$, $\tilde A^{(l)}$, $A^{(nmi)}$, 
$\tilde A^{(nml)}$ and $h^{(j)}$. 

Next, the compatibility condition of eqs.(\ref{Fz})  reads: $E^{(rsk)}_{p^{(nm)}}=E^{(nmk)}_{p^{(rs)}}$, $r,n=1,\dots,N$, $s,m=1,\dots,M$, $k=1,\dots,N+K$, or
\begin{eqnarray}\label{comp2}
\sum_{i=1}^N F^{(k)}_{u^{(i)}} S^{(rsnmi)} + \sum_{l=1}^K F^{(k)}_{w^{(l)}} \tilde S^{(rsnml)} =0,
\end{eqnarray}
where
\begin{eqnarray}
\label{c12}
S^{(rsnmi)}&=&
A^{(rsi)}_{p^{(nm)}} - A^{(nmi)}_{p^{(rs)}} +
\sum_{\tilde i=1}^N \left(
A^{(nmi)}_{u^{(\tilde i)}} A^{(rs\tilde i)} -A^{(rsi)}_{u^{(\tilde i)}} A^{(nm\tilde i)} 
\right)+\\\nonumber
&&
\sum_{\tilde l=1}^K \left(
A^{(nmi)}_{w^{(\tilde l)}} \tilde A^{(rs\tilde l)} -A^{(rsi)}_{w^{(\tilde l)}} \tilde A^{(nm\tilde l)} 
\right),
\\\label{tc12}\label{c02}
\tilde S^{(rsnml)}&=&
\tilde A^{(rsl)}_{p^{(nm)}} - \tilde A^{(nml)}_{p^{(rs)}} +
\sum_{\tilde i=1}^N \left(
\tilde A^{(nml)}_{u^{(\tilde i)}} A^{(rs\tilde i)} -\tilde A^{(rsl)}_{u^{(\tilde i)}} A^{(nm\tilde i)} 
\right)+\\\nonumber
&&
\sum_{\tilde l=1}^K \left(
\tilde A^{(nml)}_{w^{(\tilde l)}} \tilde A^{(rs\tilde l)} -\tilde A^{(rsl)}_{w^{(\tilde l)}} \tilde A^{(nm\tilde l)} 
\right).
\end{eqnarray}
 Similar to eq.(\ref{comp1}), eq.(\ref{comp2}) represents the linear relation among derivatives $F^{(k)}_{u^{(i)}}$ and  $ F^{(k)}_{w^{(l)}} $ ($\forall $ $i,l,k$). Again, condition (\ref{wr}) suggests us 
to take  all coefficients in  eq.(\ref{comp2}) equal  to zero:
\begin{eqnarray}\label{eqs_for_A2}
&&
S^{(rsnmi)}= \tilde S^{(rsnml)}=0, \\\nonumber
&&
 i,r,n=1,\dots,N,\;\; s,m=1,\dots,M,\;\; l=1,\dots,K.
\end{eqnarray}
Equations (\ref{eqs_for_A2}) represent the second system of nonlinear PDEs defining the functions $A^{(nmi)}$ and
$\tilde A^{(nml)}$. Thus, if the functions $A^{(i)}$,
$\tilde A^{(l)}$, $A^{(nmi)}$,
$\tilde A^{(nml)}$ and $h^{(j)}$ satisfy the system of nonlinear PDEs (\ref{eqs_for_A1},\ref{eqs_for_A12},\ref{eqs_for_A2}), then homogeneous system of linear PDEs (\ref{Ft}) and (\ref{Fz}) is compatible.

\subsubsection{Integration of  system (\ref{eqs_for_A1},\ref{eqs_for_A12},\ref{eqs_for_A2})}
Hereafter we consider such potentials in eqs.(\ref{Ft},\ref{Fz}) which do not  depend on $x$ explicitly. 
Then the terms with $x_j$-derivatives disappear from eqs.(\ref{c11},\ref{tc11}).
In order to write  the system of nonlinear PDEs (\ref{nleq1},\ref{nleq2}) in 
the explicit form  we have to find the admissible  expressions for the coefficients in the linear system (\ref{Ft},\ref{Fz}), i.e. we have to solve the system (\ref{eqs_for_A1},\ref{eqs_for_A12},\ref{eqs_for_A2}), which is nonlinear in general.
In the simplest cases (see  Sec.\ref{Section:ex_appl}), this system  reduces to the linear one with constant coefficients which may be integrated directly. 

In general, the following theorem provides us a big manifold of acceptable 
functions $A^{(i)}$, $\tilde A^{(j)}$, $A^{(nmi)}$ and $\tilde A^{(nmj)}$.

{\bf Theorem 2.}
Let $\Psi^{(k)}$, $k=1,\dots,N+M$, be  arbitrary functions  of variables $t$,  $u$, $p$, $w$ such that the system 
\begin{eqnarray}\label{Psit}
H^{(k)}&:=&D_t\Psi^{(k)} = \sum_{i=1}^N D_{u^{(i)}}\Psi^{(k)} A^{(i)} + 
 \sum_{l=1}^K D_{w^{(l)}}\Psi^{(k)} \tilde A^{(l)} ,\\\label{Psiz}
H^{(nmk)}&:=&D_{p^{(nm)}}\Psi^{(k)}= \sum_{i=1}^N D_{u^{(i)}}\Psi^{(k)}A^{(nmi)} + 
\sum_{l=1}^K D_{w^{(l)}} \Psi^{(k)} \tilde A^{(nml)},\\\nonumber
&& 
n=1,\dots,N,\;\;m=1,\dots,M,\;\;k=1,\dots,N+K
\end{eqnarray}
can be viewed as a  uniquely solvable system of algebraic equations for  the functions $A^{(i)}$, $\tilde A^{(l)}$, $A^{(nmi)}$ and $\tilde A^{(nml)}$, i.e.
\begin{eqnarray}\label{det}
\left|
\begin{array}{cccccc}
D_{u^{(1)}}\Psi^{(1)} & \cdots &D_{u^{(N)}}\Psi^{(1)}&D_{w^{(1)}}\Psi^{(1)}& \cdots &D_{w^{(K)}}\Psi^{(1)}\cr
\cdots&\cdots&\cdots&\cdots&\cdots&\cdots\cr
D_{u^{(1)}}\Psi^{(N+K)} & \cdots &D_{u^{(N)}}\Psi^{(N+K)}&D_{w^{(1)}}\Psi^{(N+K)}& \cdots &D_{w^{(K)}}\Psi^{(N+K)}
\end{array}
\right|\neq 0.
\end{eqnarray}
Here  we introduce operators
\begin{eqnarray}
D_y\Psi^{(k)} = \Psi^{(k)}_y + \lambda^{(y;k)} \Psi^{(k)},
\end{eqnarray}
where  $y$ is one of the variables $t$, $u^{(i)}$, $w^{(l)}$, $p^{(ij)}$. Then functions 
$A^{(i)}$, $\tilde A^{(l)}$, $A^{(nmi)}$ and $\tilde A^{(nml)}$ satisfy the system (\ref{eqs_for_A1},\ref{eqs_for_A2}).

{\it Proof.} To prove this theorem one has to show that  system 
 (\ref{eqs_for_A1},\ref{eqs_for_A2}) is the compatibility condition of the overdetermined system (\ref{Psit},\ref{Psiz}).
 In fact, the compatibility condition of eqs.(\ref{Psit}) and 
(\ref{Psiz}) reads: $ D_{p^{(nm)}}H^{(k)} = D_t H^{(nmk)}$, $\forall \;n,m,k$, or
\begin{eqnarray}\label{comp1Psi}
&&
\sum_{i=1}^N \Psi^{(k)}_{u^{(i)}} Q^{(nmi)}_0 + \sum_{l=1}^K \Psi^{(k)}_{w^{(l)}} \tilde Q^{(nml)}_0 =0,\\\nonumber
&&
n=1,\dots,N,\;\;m=1,\dots,M,\;\;k=1,\dots, N+K,
\end{eqnarray}
where 
$Q^{(nmi)}_0$ and $Q^{(nml)}_0$ equal to $Q^{(nmi)}$ and $Q^{(nml)}$   respectively  without terms having derivatives with respect to $x$, see eqs.(\ref{c11},\ref{tc11}).

Next, the compatibility conditions of eqs.(\ref{Psiz}) read: $ D_{p^{(nm)}}H^{(rsk)} = D_{p^{(rs)}}H^{(nmk)}$, $\forall \; n,m,r,s,k$, or
\begin{eqnarray}\label{comp2Psi}
&&
\sum_{i=1}^N \Psi^{(k)}_{u^{(i)}} S^{(rsnmi)} + \sum_{l=1}^K \Psi^{(k)}_{w^{(l)}} \tilde S^{(rsnml)}=0,\\\nonumber
&&
r,n=1,\dots,N,\;\;s,m=1,\dots,M,\;\;k=1,\dots, N+K,
\end{eqnarray}
where coefficients $S^{(rsnmi)}$ and $\tilde S^{(rsnml)}$ are give by eqs.(\ref{c12},\ref{c02}).
Both eqs.(\ref{comp1Psi}) and (\ref{comp2Psi})  may be viewed as  linear homogeneous systems for the functions
$Q^{(nmi)}_0$, $Q^{(nml)}_0$ and $S^{(rsnmi)}$,  $S^{(rsnmk)}$ respectively. However, due to condition (\ref{det}), the homogeneous systems (\ref{comp1Psi}) and (\ref{comp2Psi}) have only  zero solutions, which yield the systems  (\ref{eqs_for_A1},\ref{eqs_for_A2}).
$\Box$

In turn, the  functions $h^{(j)}$, $j=1,\dots,M$, may be found as solutions to the linear PDEs (\ref{eqs_for_A12}).

{\bf Remark:}
Operators $D_y$ are introduced in order to provide us the possibility to remove a dependency on some particular variable. For instance, if we would like to eliminate explicit dependency of functions  $A^{(i)}$, $\tilde A^{(l)}$, $A^{(nmi)}$, $\tilde A^{(nml)}$ and $h^{(j)}$ on some particular variable $y_0$, then we must remove  derivative in the appropriate operator $D_{y_0}$ which acquires  the following form: $D_{y_0}\Psi^{(k)} =  \lambda^{(y_0;k)} \Psi^{(k)}$.


\subsubsection{Admissible reductions for the potentials of eqs.(\ref{Ft}) and (\ref{Fz})}
The described method for construction of potentials in eqs.(\ref{Ft}) and (\ref{Fz}) preserves $(N+K)$ arbitrary  functions $\Psi^{(k)}$ of all variables $t$, $u$, $w$, $p$. The availability of such freedom  suggests us to look for additional compatible relations among the above potentials. A big class of these relations may be established by means of the spectral theory developed for the nonlinear PDEs assotiated with commuting vector fields \cite{MS1,MS2}. Namely, the following theorem is valid.

{\bf Theorem 3.}
Let us consider the  spectral problem for some spectral function $\chi(\lambda)$:
\begin{eqnarray}\label{Psit2}
{\cal{H}}^{(t)}&:=&\chi_t(\lambda)+\lambda\chi_{\eta_{00}}(\lambda)  = \sum_{i=1}^N D_{u^{(i)}}\chi(\lambda) A^{(i)} + 
 \sum_{l=1}^K D_{w^{(l)}}\chi(\lambda) \tilde A^{(l)},\\\label{Psiz2}
{\cal{H}}^{(nm)}&:=&\chi_{p^{(nm)}}(\lambda)+\lambda\chi_{\eta_{nm}}(\lambda)= \sum_{i=1}^N D_{u^{(i)}}\chi(\lambda) A^{(nmi)} + \\\nonumber
&&
\sum_{l=1}^K D_{w^{(l)}}\chi(\lambda) \tilde A^{(nml)} ,\;\;n=1,\dots,N,\;\;m=1,\dots,M,
\end{eqnarray}
where $\eta_{ij}$ are additional parameters, which, in particular, may coincide with some parameters $t$, $u$, $w$ or $p$; $\lambda$ is a complex spectral parameter. 
The spectral function $\chi$ depends on $\lambda$ as well as  on all other parameters such as $t$, $u$, $p$, $w$ and $\eta_{ij}$. 
Then functions $A^{(i)}$, $\tilde A^{(l)}$, $A^{(nmi)}$, $\tilde A^{(nml)}$ and $h^{(j)}$ satisfy the system (\ref{eqs_for_A1}) (without terms having $x_j$-derivatives)
and (\ref{eqs_for_A2})  as well as additional reductions
\begin{eqnarray}
\label{chired}
&&
A^{(i)}_{\eta_{nm}} =A^{(nmi)}_{t},\;\;
\tilde A^{(l)}_{\eta_{nm}} =\tilde A^{(nml)}_{t},\;\;A^{(rsi)}_{\eta_{nm}} =A^{(nmi)}_{p^{(rs)}},\;\;\tilde A^{(rsl)}_{\eta_{nm}} =\tilde A^{(nml)}_{p^{(rs)}},\\\nonumber
&&
n,r,i=1,\dots,N,\;\;s,m=1,\dots,M,\;\;l=1,\dots,K.
\end{eqnarray} 

{\it Proof.}
Overdetermined linear system (\ref{Psit2},\ref{Psiz2}) for the function $\chi$ must be compatible. Compatibility conditions of eqs.(\ref{Psit2}) and (\ref{Psiz2}) read:
${\cal{H}}^{(t)}_{p^{(nm)}}={\cal{H}}^{(nm)}_t$, $\forall$ $n,m$, or
\begin{eqnarray}\label{comp1chi}
&&
\sum_{i=1}^N D_{u^{(i)}} \chi^{(k)} Q^{(nmi)}_0+ \sum_{l=1}^K D_{w^{(l)}}\chi^{(k)} \tilde Q^{(nml)}_0+\\\nonumber
&&
\lambda\left(\sum_{i=1}^N D_{u^{(i)}}\chi^{(k)} (A^{(i)}_{\eta_{nm}} -A^{(nmi)}_{t}) +\sum_{l=1}^KD_{w^{(l)}} \chi^{(k)}(\tilde A^{(l)}_{\eta_{nm}} -\tilde A^{(nml)}_{t})   \right)
=0,\\\nonumber
&&
n=1,\dots,N,\;\;m=1,\dots,M,\;\;k=1,\dots, N+K.
\end{eqnarray}
Next, the compatibility conditions of eqs.(\ref{Psiz2}) read: $ D_{p^{(nm)}}H^{(rs)} = D_{p^{(rs)}}H^{(nm)}$, $\forall n,m,r,s$, or
\begin{eqnarray}\label{comp2chi}
&&
\sum_{i=1}^N D_{u^{(i)}}\chi^{(k)} S^{(rsnmi)} + \sum_{l=1}^K D_{w^{(l)}}\chi^{(k)} \tilde S^{(rsnml)}
+\\\nonumber
&&
\lambda\left(\sum_{i=1}^N D_{u^{(i)}}\chi^{(k)} (A^{(rsi)}_{\eta_{nm}} -
A^{(nmi)}_{p^{(rs)}}) + 
\sum_{l=1}^K D_{w^{(l)}}\chi^{(k)}(\tilde A^{(rsl)}_{\eta_{nm}} -\tilde A^{(nml)}_{p^{(rs)}})  \right)
=0,\\\nonumber
&&
r,n=1,\dots,N,\;\;s,m=1,\dots,M,\;\;k=1,\dots, N+K.
\end{eqnarray}
Following the general strategy of spectral theory for the equations associated with commuting vector fields \cite{MS1,MS2} we conclude that all coefficients in eqs.(\ref{comp1chi},\ref{comp2chi}) are zeros, which yield relations (\ref{chired}) along with eqs.(\ref{eqs_for_A1},\ref{eqs_for_A2}). $\Box$

Again, the appropriate functions $h^{(j)}$ may be found as solutions to the linear system (\ref{eqs_for_A12}).

{\bf Remark:} Relations (\ref{chired})
may be satisfied, if we introduce functions ${\cal{A}}^{(i)}$ and $\tilde {\cal{A}}^{(k)}$ by the following formulas:
\begin{eqnarray}\label{red}
A^{(i)}={\cal{A}}^{(i)}_{\eta_{00}},\;\;\tilde A^{(l)}=\tilde {\cal{A}}^{(l)}_{\eta_{00}},\;\;
A^{(nmi)}={\cal{A}}^{(i)}_{\eta_{nm}},\;\;\tilde A^{(nml)}=\tilde {\cal{A}}^{(l)}_{\eta_{nm}}.
\end{eqnarray}


\section{Simple examples of nonlinear PDEs and implicit representation  of  their solutions}
\label{Section:examples}
After functions $A^{(i)}$, $\tilde A^{(l)}$, $A^{(nmi)}$, $\tilde A^{(nml)}$ and $h^{(j)}$ have been constructed, one has to find $(N+K)$ solutions $F^{(k)}$, $k=1,\dots,N+K$, to the linear system (\ref{Ft},\ref{Fz}) such that condition (\ref{wr}) is satisfied. This problem may be solved numerically, in general. 
The last step in construction of solutions to nonlinear PDEs (\ref{nleq1},\ref{nleq2}) is solving the system of  $M$-dimensional  PDEs (\ref{F}) for the functions $u$ and $w$.
Some examples of nonlinear PDEs together with implicit description of their solutions are represented below.

\subsection{ Example 1:  the system of first order nonlinear PDEs with the same leading part} 
\label{Section:ex_cl}
We show that our algorithm reduces to the method of  characteristics in the particular case.
Let $A^{(nmi)}=\tilde A^{(nml)}=0$ and 
$F^{(k)}$ be independent on $p$. Then eq.(\ref{Fz}) disappears, so that the functions $F^{(k)}$ become  solutions to the single linear PDE (\ref{Ft}):
\begin{eqnarray}\label{Ft_cl}
F^{(k)}_t +\sum_{j=1}^M F^{(k)}_{x_j} h^{(j)} = \sum_{i=1}^N F^{(k)}_{u^{(i)}} A^{(i)} + 
 \sum_{l=1}^K F^{(k)}_{w^{(l)}} \tilde A^{(l)},\;\;k=1,\dots,N+K
\end{eqnarray}
with arbitrary functions  $A^{(i)}$, $\tilde A^{(j)}$ and $h^{(j)}$ of all variables $t$, $x$, $u$  and $w$. Eq.(\ref{Ft_cl}) is equivalent to eq.(\ref{F_lin_cl}).
In turn,  nonlinear PDEs (\ref{nleq1},\ref{nleq2}) read
\begin{eqnarray}\label{nleq_cl}
&&
u^{(i)}_t+\sum_{n=1}^M u^{(i)}_{x_n}h^{(n)} + A^{(i)}=0,\\\label{nleq2_cl}
&&
w^{(i)}_{t}+\sum_{n=1}^M w^{(i)}_{x_n} h^{(n)}+ \tilde A^{(i)}=0.
\end{eqnarray}
This is a system of nonlinear PDEs with the same leading part, which is analogous to  system (\ref{cl_lead}). We do not discuss its solutions.

\subsection{ Example 2: nonlinear PDEs in the form $\Big(t+f^{(k)}(u,p)\Big)_t=0$, $k=1,\dots,N$}
\label{Section:ex_expl}
In this subsection we consider an example of the system of nonlinear PDEs which may be given the following general form: 
\begin{eqnarray}
\label{sf}
\Big(t+f^{(k)}(u,p)\Big)_t=0,\;\;k=1,\dots,N.
\end{eqnarray}
 Of course, this system may be immediately integrated over $t$ yielding $t+f^{(k)}(u,p)=c^{(k)}(x)$, where $c^{(k)}$ are arbitrary functions of $x$, so that we do not need to apply our algorithm. However, we represent derivation of such equations using our algorithm in order to demonstrate its relation with the classical integration methods.

Let $K=0$ and $\Psi^{(k)}$ be the following linear functions of $u$ and $p$ (independent on $w$):
\begin{eqnarray}
\Psi^{(k)}=  u^{(k)}  + 
 \sum_{i=1}^N \sum_{j=1}^M \alpha^{(ijk)}p^{(ij)},\;\;k=1,\dots,N,\;\;,\;\;
\alpha^{(ijk)}=const.
\end{eqnarray}
We take
\begin{eqnarray}
&&
D_t\Psi^{(k)}=\lambda^{(k)}\Psi^{(k)},\;\;\lambda^{(u^{(n)})}=0,\;\;
\lambda^{(p^{(nm)};k)}=\lambda^{(nmk)},\;\;h^{(m)}=0,\\\nonumber
&&
k,n=1,\dots,N,\;\;m=1,\dots,M.
\end{eqnarray}
Algebraic system (\ref{Psit},\ref{Psiz}) becomes simpler in this case:
\begin{eqnarray}\label{Psit_ex3}
&&
\lambda^{(k)}\Psi^{(k)} = \sum_{i=1}^N \Psi^{(k)}_{u^{(i)}} A^{(i)} ,\\\label{Psiz_ex3}
&&
D_{p^{(nm)}}\Psi^{(k)}= \sum_{i=1}^N \Psi^{(k)}_{u^{(i)}}A^{(nmi)},\;\;k,n=1,\dots,N,\;\;m=1,\dots,M.
\end{eqnarray}
Its solution reads:
\begin{eqnarray}\label{ex2_A}
&&
A^{(k)}=\lambda^{(k)}\Psi^{(k)}
, \;\;\; A^{(nmk)}=
\alpha^{(nmk)} +   \lambda^{(nmk)}\Psi^{(k)},\;\;
k,n=1,\dots,N,\;\;m=1,\dots,M.
\end{eqnarray}
Regarding the nonlinear system, we observe that eq.(\ref{nleq2}) disappears while eq.(\ref{nleq1}) reads:
\begin{eqnarray}\label{red1_u_ex}
&&
u^{(i)}_t + A^{(i)}+\sum_{j=1}^N \sum_{m=1}^M A^{(jmi)} 
u^{(j)}_{x_m t} =0,\;\;k,n=1,\dots,N,\;\;m=1,\dots,M.
\end{eqnarray}

Now we consider the system (\ref{Ft},\ref{Fz}) defining functions $F^{(k)}$, $k=1,\dots,N$.
Introducing  new variables $\Psi^{(i)}$ instead of $u^{(i)}$, $i=1,\dots,N$,
we  rewrite this system  as follows:
\begin{eqnarray}\label{F_ex3}
F^{(k)}_t=\sum_{i=1}^N F^{(k)}_{\Psi^{i}}\Psi^{i} \lambda^{(i)} ,\;\;\;
 F^{(k)}_{p^{(nm)}}=\sum_{i=1}^N F^{(k)}_{\Psi^{i}}\Psi^{i} \lambda^{(nmi)},
\;\;k,n=1,\dots,N,\;\;m=1,\dots,M.
\end{eqnarray}
Solution to  system (\ref{F_ex3})  reads:
\begin{eqnarray}\label{FG}
F^{(k)}&=&G^{(k)}\Big(x_1,\dots,x_M,\ln\Psi^{(1)} + \lambda^{(1)} t + \sum_{i=1}^N \sum_{j=1}^M p^{(ij)} \lambda^{(ij1)}, \dots,\\\nonumber
&&
 \ln\Psi^{(N)} + \lambda^{(N)} t + \sum_{i=1}^N \sum_{j=1}^M p^{(ij)} \lambda^{(ijN)}
\Big), \;\;
k=1,\dots,N,
\end{eqnarray}
where  $G^{(k)}$ are arbitrary functions of $M+N$ variables.
We may choose functions  $G^{(k)}$ such that  the system  
$G^{(k)}(x_1,\dots,x_M,z_1,\dots,z_N)=0$, $k=1,\dots,N$, (see eqs.(\ref{F})) is uniquely solvable with respect to variables $z_1,\dots,z_N$. Then system (\ref{F}) may be replaced with the following one:
\begin{eqnarray}\label{rho}
\ln \Psi^{(k)} + \lambda^{(k)} t + \sum_{i=1}^N \sum_{j=1}^M p^{(ij)} \lambda^{(ijk)}=C^{(k)}(x_1,\dots,x_M),\;\;k=1,\dots,N,
\end{eqnarray}
where $C^{(i)}$ are  arbitrary functions of $M$ variables.
It is obvious that the $t$-derivatives of eqs.(\ref{rho}) yield the system (\ref{red1_u_ex}), so that we deal with the nonlinear PDEs writable in  form (\ref{sf}).
Eqs.(\ref{rho}) are  $M$-dimensional nonlinear PDEs where $t$ appears as a parameter.

The above arbitrary functions $C^{(k)}$ may be used in order to solve the initial value problem. In fact 
let $u_0=u|_{t=0}$ be the initial condition. The appropriate initial values  $p^{(ij)}_0$ and $\Psi^{(i)}_0$ for the functions $p^{(ij)}\equiv u^{(i)}_{x_j}$ and $\Psi^{(i)}$  may be calculated directly. Putting $t=0$ in eqs.(\ref{rho}) one gets the following equations:
 \begin{eqnarray}\label{rho0}
\ln \Psi^{(k)}_0  + \sum_{i=1}^N\sum_{j=1}^M p^{(ij)}_0 \lambda^{(ijk)}=C^{(k)}(x_1,\dots,x_M),\;\;k=1,\dots,N,
\end{eqnarray}
which uniquely define functions  $C^{(k)}$, $k=1,\dots,N$.

In particular, let $N=1$, $M=2$, $u\equiv u^{(1)}$, $\lambda\equiv \lambda^{(1)}$, so that $\Psi^{(1)}=u+\alpha^{(111)} u_{x_1} +\alpha^{(121)} u_{x_2}$ and    eq.(\ref{red1_u_ex}) reduces to the single nonlinear PDE:
\begin{eqnarray}\label{nl2}
&&
u_t+\lambda (u_{\xi_1}+ u)+u_{\xi_1 t} + (u_{\xi_1}+u)u_{\xi_2 t}=0,
\end{eqnarray}
where we introduce the new variables $\xi_i$, $i=1,2$ in accordance with the following formulas:
$\partial_{\xi_1}=\alpha^{(111)}\partial_{x_1} + \alpha^{(121)} \partial_{x_2}$,
$\partial_{\xi_2}=\lambda^{(111)} \partial_{x_1} + \lambda^{(121)} \partial_{x_2}$.
The system (\ref{rho}) reduces to the single two-dimensional PDE, where $t$ appears as a parameter: 
\begin{eqnarray}\label{Fex2}
\ln (u_{\xi_1} + u) +\lambda t + u_{\xi_2}=C^{(1)}(\xi_1,\xi_2).
\end{eqnarray}
If there is no dependency on  $\xi_1$, i.e. $u_{\xi_1}=0$, then  eq.(\ref{nl2}) reads
\begin{eqnarray}\label{ex3}
u_t+u u_{\xi_2 t}+\lambda u =0.
\end{eqnarray}
In this case eq.(\ref{Fex2}) becomes ODE, $C^{(1)}(\xi_1,\xi_2)\to C(\xi_2)$: 
\begin{eqnarray}\label{Fex3}
\ln u +\lambda t +  u_{\xi_2} =C(\xi_2),
\end{eqnarray}
where $C(\xi_2)$ is an arbitrary function of single variable.

\subsection{Example 3: a new class of nonlinear PDEs
} 
\label{Section:ex_appl}
\label{Section:example_hydr}

In this section we consider an example when equations  (\ref{eqs_for_A1}), (\ref{eqs_for_A12}) and
(\ref{eqs_for_A2}) reduce to the linear ones with constant coefficients and consequently may be easily integrated.
In addition, we assume that all functions appearing in system of equations (\ref{eqs_for_A1}), (\ref{eqs_for_A12}) and
(\ref{eqs_for_A2}) do not depend on $t$, $x_j$  and $u_i$, so that the appropriate derivatives disappear from these equations.

\subsubsection{The general solution to eqs.  (\ref{eqs_for_A1}), (\ref{eqs_for_A12}) and
(\ref{eqs_for_A2}) with $\tilde A^{(i)}=const$ and $\tilde A^{(nmj)}=const$ and appropriate nonlinear PDEs.}
\label{Section:lin_system}

Let $\tilde A^{(nmi)}=\gamma^{(nmi)}=const$ and 
$\tilde A^{(i)}= \gamma^{(i)}=const$.  Then equations (\ref{eqs_for_A1}), (\ref{eqs_for_A12}) and
(\ref{eqs_for_A2}) reduce to the following system:
\begin{eqnarray}\label{ex2:At}
&&
A^{(i)}_{p^{(nm)}}   +
\sum_{\tilde l=1}^K \left(
A^{(nmi)}_{w^{(\tilde l)}} \gamma^{(\tilde l)} -A^{(i)}_{w^{(\tilde l)}} \gamma^{(nm\tilde l)} 
\right)=0,
\\\label{ex2:h}
&&
h^{(j)}_{p^{(nm)}} 
-
\sum_{\tilde l=1}^K h^{(j)}_{w^{(\tilde l)}} \gamma^{(nm\tilde l)}=0,
\\\label{ex2:Ap}
&&
A^{(rsi)}_{p^{(nm)}} - A^{(nmi)}_{p^{(rs)}} +
\sum_{\tilde l=1}^K \left(
A^{(nmi)}_{w^{(\tilde l)}} \gamma^{(rs\tilde l)} -A^{(rsi)}_{w^{(\tilde l)}} \gamma^{(nm\tilde l)} 
\right)=0.
\end{eqnarray}
Solution to eq.(\ref{ex2:h}) reads:
\begin{eqnarray}
&&
h^{(j)}=g^{(j)}\left(\xi\right),\;\;\;j=1,\dots,K,\\\nonumber
&&
\xi=\left\{\xi^{(i)}=w^{(i)} +\sum_{n=1}^N\sum_{m=1}^M \gamma^{(nmi)}p^{(nm)}:i=1,\dots,K\right\}.
\end{eqnarray}
System (\ref{ex2:At},\ref{ex2:Ap})   may be readily integrated.  Assuming that (at least)  some $\gamma^{(i)}$ are nonzero one obtains:
\begin{eqnarray}\label{ex2_expl_At}
&&
A^{(i)}=\int_\Gamma d\Omega(k) a^{(i)}(k) e^{i\left(\sum\limits_{l_1=1}^K k^{(l_1)} w^{(l_1)} + 
\sum\limits_{i_1=1}^N \sum\limits_{j_1=1}^M k^{(i_1j_1)} p^{(i_1j_1)}\right)}+G^{(i)}(\xi),\\\label{ex2_expl_Ap}
&&
A^{(nmi)}=\int_\Gamma  d\Omega(k) a^{(nmi)}(k) e^{i\left(\sum\limits_{l_1=1}^K k^{(l_1)} w^{(l_1)} + 
\sum\limits_{i_1=1}^N \sum\limits_{j_1=1}^M k^{(i_1j_1)} p^{(i_1j_1)}\right)}+G^{(nmi)}_0,\\\label{ex2_expl_aa}
&&
a^{(nmi)}(k)=\frac{a^{(i)}(k)
\left(\sum\limits_{l=1}^K k^{(l)} \gamma^{(nm l)}-k^{(nm)} \right)
}{\sum\limits_{l=1}^K k^{(l)} \gamma^{(l)} },
\\\nonumber
&&
i,n=1,\dots,N,\;\;m=1,\dots,M,\\\nonumber
&&
k=\{
k^{(i)}: i=1,\dots,K; k^{(nm)}: n=1,\dots,N,\;m=1,\dots,M 
\},
\end{eqnarray}
where $\omega(k)$ is some measure and $\Gamma$ is a  domain in the space of vector parameter $k$; $G^{(nmi)}_0=const$, $G^{(i)}(\xi)$ are arbitrary functions of $K$ variables $\xi^{(i)}$, $i=1,\dots,K$.
To satisfy eq.(\ref{ex2_expl_aa})  we take the following structure of functions $a^{(nmi)}(k)$ 
and $a^{(i)}(k)$:
\begin{eqnarray}\label{anmi}
&&
a^{(nmi)}(k)= b^{(i)}(k)\left(\sum_{l=1}^K \gamma^{(nml)} k^{(l)}-k^{(nm)}\right),\\\label{ai}
&&
a^{(i)}(k)=b^{(i)}(k)\sum\limits_{l=1}^K k^{(l)} \gamma^{(l)},
\end{eqnarray} 
where $b^{(i)}(k)$, $i=1,\dots,N$, are arbitrary functions.

In particular, if all $\gamma^{(i)}=0$, then $a^{(i)}(k)=0$ and eq.(\ref{ex2_expl_At}) reduces to the following one:
\begin{eqnarray}
&&
A^{(i)}=G^{(i)}(\xi),
\end{eqnarray}
while eq.(\ref{ex2_expl_aa}) must be replaced with   the following relations among $
a^{(nmi)}$:
\begin{eqnarray}
&&
a^{(rsi)}(k)=a^{(nmi)}(k) \frac{
\sum_{l=1}^K \gamma^{(rsl)} k^{(l)}-k^{(rs)}}{\sum_{l=1}^K \gamma^{(nml)} k^{(l)}-k^{(nm)}},
\end{eqnarray}
which agrees with eqs.(\ref{anmi}).
Nonlinear system (\ref{nleq1},\ref{nleq2}) reads as follows:
\begin{eqnarray}\label{ex2:nleq1}
&&
u^{(i)}_t+\sum_{n=1}^M u^{(i)}_{x_n}g^{(n)}(\xi) + A^{(i)}+\sum_{j=1}^N \sum_{m=1}^M A^{(jmi)} 
\left(u^{(j)}_{x_m t} + \sum_{n=1}^M u^{(j)}_{x_m x_n} g^{(n)}(\xi)\right)=0,\\\label{ex2:nleq2}
&&
\xi^{(l)}_{t}+\sum_{n=1}^M \xi^{(l)}_{x_n} g^{(n)}(\xi)+ \gamma^{(l)}=0,\;\;\;
i=1,\dots,N,\;\;l=1,\dots,K.
\end{eqnarray}
Finally, eqs.(\ref{Ft},\ref{Fz}) read:
\begin{eqnarray}\label{FtAp_const}
&&
F^{(k)}_t +\sum_{j=1}^M F^{(k)}_{x_j} h^{(j)} = \sum_{i=1}^N F^{(k)}_{u^{(i)}} A^{(i)} + 
 \sum_{l=1}^K F^{(k)}_{w^{(l)}} \gamma^{(l)} ,\\\label{FzAp_const}
&&
F^{(k)}_{p^{(nm)}}= \sum_{i=1}^N F^{(k)}_{u^{(i)}} A^{(nmi)} + 
\sum_{l=1}^K F^{(k)}_{w^{(l)}} \gamma^{(nml)},\\\nonumber
&&
n=1,\dots,N,\;\;m=1,\dots,M,\;\;k=1,\dots,N+K .
\end{eqnarray}
Solutions $F^{(k)}$ to  system (\ref{FtAp_const},\ref{FzAp_const}) must be used in eqs.(\ref{F}) for construction the functions $u$ and $w$.

\subsubsection{An application to the hydrodynamics of viscous fluid}
Now we consider a simple  particular solution to the system (\ref{ex2:At}-\ref{ex2:Ap}) together with the appropriate nonlinear PDEs in more details. This case  agrees with the general solution derived in Sec.\ref{Section:lin_system}. 
Let us take  $N=K=1$, $M=2$, $u^{(1)}\equiv u$, $w^{(1)}\equiv w$, 
$p^{(nm)}\equiv p^{(m)}=u_{x_m}$, $h^{(i)}\equiv 0$.
Then eqs.(\ref{ex2:At}-\ref{ex2:Ap}) reduce to the following ones:
\begin{eqnarray}\label{Rex2:At}
&&
A^{(1)}_{p^{(m)}}   +
A^{(1m1)}_{w} \gamma^{(1)}-A^{(1)}_w\gamma^{(1m1)}=0,\;\;m=1,2,
\\\label{Rex2:Ap}
&&
A^{(1s1)}_{p^{(m)}} - A^{(1m1)}_{p^{(s)}} + A^{(1m1)}_w \gamma^{(1s1)} - A^{(1s1)}_w \gamma^{(1m1)}=0,\;\;s,m=1,2.
\end{eqnarray}
In turn, the system of nonlinear equations
(\ref{ex2:nleq1},\ref{ex2:nleq2}) gets the following form:
\begin{eqnarray}\label{Rex2:nleq1}
&&
u_t+ A^{(1)}+\sum_{m=1}^2 A^{(1m1)} 
u_{x_m t} =0,\\\label{Rex2:nleq2}
&&
w_{t}+ \gamma^{(1)}+\sum_{m=1}^2 \gamma^{(1m1)} u_{x_mt}=0.
\end{eqnarray}
The general solution to eqs.(\ref{Rex2:At},\ref{Rex2:Ap}) reads:
\begin{eqnarray}\label{Am}
A^{(1m1)}=-\frac{1}{\gamma^{(1)}} \partial^{-1}_w A^{(1)}_{p^{(m)}} + \frac{\gamma^{(1m1)}}{\gamma^{(1)}}A^{(1)} + c^{(m)},
\end{eqnarray}
where $A^{(1)}$ is an arbitrary function of $w$, $p^{(1)}$ and $p^{(2)}$ while $c^{(m)}$ ($m=1,2$) are  arbitrary constants.
We look for the solution to  linear system (\ref{FtAp_const},\ref{FzAp_const}) in the following form:
\begin{eqnarray}
F^{(k)}=H^{(k)}(z,u,p^{(1)},p^{(2)},t) + C^{(k)}(x_1,x_2),\;\;z=w+\sum_{m=1}^2 p^{(m)}\gamma^{(1m1)} + \gamma^{(1)} t,
\end{eqnarray}
where $C^{(k)}$ ($k=1,2$) are  arbitrary functions of two variables
while $H^{(k)}$ ($k=1,2$) satisfy the following linear system:
\begin{eqnarray}\label{RFtAp_const}
&&
H^{(k)}_t  =  H^{(k)}_{u} \left(A^{(1)}|_{w=z-\sum_{m=1}^2 p^{(m)}\gamma^{(1m1)}-\gamma^{(1)} t}\right) ,\\\label{RFzAp_const}
&&
H^{(k)}_{p^{(m)}}= H^{(k)}_{u} \left(A^{(1m1)}|_{w=z-\sum_{m=1}^2 p^{(m)}\gamma^{(1m1)}-\gamma^{(1)} t}\right),\;\;
m=1,2.
\end{eqnarray}
It is simple to show that, if $A^{(1)}$ does not depend on $w$, then  system (\ref{RFtAp_const},\ref{RFzAp_const}) may be  integrated analytically and  the $t$-derivatives may be removed from  nonlinear PDEs (\ref{Rex2:nleq1},\ref{Rex2:nleq2})  by the simple  method discussed in  Sec.\ref{Section:ex_expl}. In general, one has to solve system  (\ref{RFtAp_const},\ref{RFzAp_const}) numerically, so that the appropriate nonlinear PDEs are beyond the family of PDEs considered in Sec.\ref{Section:ex_expl}. Namely this case leads to the new partially integrable nonlinear PDEs.

The final step is solving the two-dimensional first order PDE (\ref{F}), which reads as follows:
\begin{eqnarray}\label{ex_H}
H^{(k)}(z,u,p^{(1)},p^{(2)},t) + C^{(k)}(x_1,x_2)=0,\;\;k=1,2.
\end{eqnarray}
In addition, functions $F^{(k)}$ must  satisfy the condition (\ref{wr}) which now reads:
 \begin{eqnarray}\label{wrH}
\left.\frac{\partial  (H^{(1)},  H^{(2)})}{\partial (z,u)}\right|_{{\mbox{Eqs.(\ref{ex_H})}}} \neq 0.
\end{eqnarray}
Without loss of generality, we take $H^{(2)}(z,u,p^{(1)},p^{(2)},t)\equiv z$,
so that  system (\ref{ex_H}) can be written as follows:
\begin{eqnarray}\label{2ex_H}
&&
H^{(1)}(z,u,p^{(1)},p^{(2)},t) + C^{(1)}(x_1,x_2)=0,\\\label{2ex_H2}
&&
z+C^{(2)}(x_1,x_2)\equiv w+\sum_{m=1}^2 p^{(m)}\gamma^{(1m1)} + \gamma^{(1)} t
+C^{(2)}(x_1,x_2)=0
\end{eqnarray}
It is obvious that  the described algorithm preserves two  arbitrary functions $C^{(m)}(x_1,x_2)$ ($m=1,2$) of  two variables in the solution space to  nonlinear system (\ref{Rex2:nleq1},\ref{Rex2:nleq2}). These functions may be uniquely fixed by the
initial conditions at $t=0$  as follows ($u_0=u|_{t=0}$,  $w_0=w|_{t=0}$, $p^{(m)}_0=(u_0)_{x_m}$):
\begin{eqnarray}\label{ivp}
 C^{(1)}(x_1,x_2)=-H^{(1)}(z,u_0,p^{(1)}_0,p^{(2)}_0,0),\;\;C^{(2)}(x_1,x_2)=-w_0-\sum_{m=1}^2 p^{(m)}_0\gamma^{(1m1)} .
\end{eqnarray}
Thus, our algorithm is suitable for the initial value problem.

Now we turn to the more specific example. Let us take $A^{(1)}$ in the following form:
\begin{eqnarray}\label{A1}
&&
A^{(1)}=-\nu (p^{(1)}+ip^{(2)}) +\frac{1}{2} (p^{(1)}+p^{(2)})^2 +w^3
\end{eqnarray}
and use the following values for the constant parameters:
\begin{eqnarray}\label{par}
 c^{(1)}=1,\;\;\; c^{(2)}=-i,\;\;\; \gamma^{(1)}=1,\;\;\; \gamma^{(111)}=\gamma^{(121)}=0.
\end{eqnarray}
 Then eq.(\ref{Am}) yields the following expressions for $A^{(1m1)}$, $m=1,2$:
\begin{eqnarray}\label{Amm}
A^{(111)}=(\nu-p^{(1)}-p^{(2)}) w +c^{(1)} ,\;\;\;A^{(121)}=(i \nu-p^{(1)}-p^{(2)}) w+c^{(2)}.
\end{eqnarray}
Substituting eqs.(\ref{A1}-\ref{Amm}) into eqs.(\ref{Rex2:nleq1}) and (\ref{Rex2:nleq2}) one obtains the following nonlinear system:
\begin{eqnarray}\label{nl1}
&&
u_t-\nu(u_{x_1}+i u_{x_2}) +\frac{1}{2}(u_{x_1}+u_{x_2})^2 +w^3+ (u_{x_1}-i u_{x_2})_t +\\\nonumber
&&
w \Big(
\nu(u_{x_1}+i u_{x_2})_t- (u_{x_1}+u_{x_2})(u_{x_1}+ u_{x_2})_t
\Big)=0,\\\label{nl22}
&&
w_t+1=0.
\end{eqnarray}
In order to construct the general solution to system (\ref{nl1},\ref{nl22}), one has to substitute eqs.(\ref{A1}-\ref{Amm}) into the system 
(\ref{RFtAp_const},\ref{RFzAp_const}) (see eqs.(\ref{intr_Hd})) and solve it for $H^{(k)}$, $k=1,2$. After that, the initial value problem  for eqs. (\ref{nl1},\ref{nl22}) may be solved by means of eqs.(\ref{2ex_H}-\ref{ivp}).
System (\ref{nl1},\ref{nl22}) is represented in the Introduction, see eqs.(\ref{intr_nl1}).

\paragraph{Small scale expansion of eqs.(\ref{nl1},\ref{nl22}) and (2+1)-dimensional dynamics of viscous fluid.}
Let us apply the differential operator $(\partial_{x_1}+\partial_{x_2})$ to the nonlinear PDE (\ref{nl1}) and 
introduce a small parameter $\varepsilon$ by the following formulas:
\begin{eqnarray}\label{small_par}
u_{x_1}+u_{x_2}=\varepsilon v,\;\;v_{x_i}=\varepsilon v_{\xi_i},\;\;w=\varepsilon^2\tilde w,\;\;v_t=\varepsilon v_\eta,
\end{eqnarray}
where $\eta$, $\xi_1$ and $\xi_2$ are the new independent variables and $v$, $\tilde w$
 are the new fields.
One obtains
\begin{eqnarray}\label{small}
\varepsilon^2\Big(v_\eta-\nu(v_{\xi_1}+i v_{\xi_2})\Big) +\varepsilon^3\Big(v (v_{\xi_1}+v_{\xi_2})+v_{\xi_1\eta}-i v_{\xi_2\eta}\Big)=o(\varepsilon^3).
\end{eqnarray}
 We see that the field $ \tilde w $ does not appear in two leading terms of eq.(\ref{small}).

Now we eliminate $\eta$-derivative in the last term of the RHS of eq.(\ref{small}) substituting 
$v_\eta=\nu(v_{\xi_1}+i v_{\xi_2}) + o(\varepsilon)$. Finally,  let us introduce the independent variables $\tau$, $y_1$ and $y_2$ instead of $\eta$, $\xi_1$ and $\xi_2$  by means of the following relations:
$v_\eta-\nu(v_{\xi_1}+i v_{\xi_2})=\varepsilon v_\tau$, $v_{\xi_i}=v_{y_i}$, $i=1,2$. Then the leading term of  eq.(\ref{small}) yields the following nonlinear PDE for the function $v$:
\begin{eqnarray}\label{res}
v_\tau+ v(v_{y_1}+v_{y_2}) - \nu (v_{y_1y_1}+v_{y_2y_2})=0,
\end{eqnarray}
which describes, for instance, the two-dimensional  dynamics of  viscous fluid with equal components of velocity. Here $\nu$ is the constant  kinematic viscosity. This equation is written in the Introduction, see eq.(\ref{intr}).

{\bf Remark 1:} The small parameter must be introduced not only in eq.(\ref{nl1}), but also in eq.(\ref{nl22}). Introducing $\varepsilon$ in the solution to eq.(\ref{nl22}) (see eq.(\ref{2ex_H2}) with constant parameters given by formulas (\ref{par})) we obtain
\begin{equation}\label{C}
w\equiv -C^{(2)}\Big(\frac{y_1}{\varepsilon}-\frac{\nu \tau}{\varepsilon^2},\frac{y_2}{\varepsilon}-\frac{i\nu \tau}{\varepsilon^2}\Big)  -\frac{\tau}{\varepsilon^2} =\varepsilon^2 \tilde w.
\end{equation}
This formula  requires the special form of the arbitrary function $C^{(2)}$.
But, as far as the field $\tilde w$ does not appear in eq.(\ref{small}) as well as in the final nonlinear PDE (\ref{res}), we do not consider eq.(\ref{C}) in more details.

{\bf Remark 2:} It is important to note that  equation (\ref{res}) may be derived as a small scale expansion of a big family of
more complicated  nonlinear PDEs inside of the class of partially integrable PDEs. The system (\ref{nl1},\ref{nl22}) is a representative of this family produced by our choice of the  function $A^{(1)}$ and constant parameters, see eqs.(\ref{A1}) and (\ref{par}). Another representative is produced by the following choice of the function $A^{(1)}$:
\begin{eqnarray}
A^{(1)} = -\nu (p^{(1)}+i p^{(2)}) +\frac{1}{w}\Big(e^{w\Big(1+\frac{1}{2}(p^{(1)}+p^{(2)})^2\Big)}-1\Big)
\end{eqnarray}
with constant  parameters given by formulas (\ref{par}). Thus, eq.(\ref{res}) is, in some sense, ''an attractor'' in a big family of partially integrable nonlinear PDEs.

\section{Conclusions}

\label{Section:conclusions}

We suggest an algorithm allowing one to reduce a big class of systems of  ($M+1$)-dimensional scalar nonlinear PDEs   to the appropriate systems of nonlinear $M$-dimensional PDEs. In particular, if $M=1$, then the $M$-dimensional  system becomes the system of ODEs. We refer to these PDEs as partially integrable ones since we are not able to integrate them over all variables. Among examples, there is a family of (2+1)-dimensional PDEs having the same  small scale limit applicable to the hydrodynamics, see eq.(\ref{res}). Existence of this example suggests us to look for further improvements of the represented  algorithm  in order to  integrate eq.(\ref{res}) directly and exhibit new   nonlinear PDEs which are both applicable and integrable. 
The represented method may be viewed as a generalization of the method of characteristics.

This work is supported by the RFBR grants  10-01-00787 and 09-01-92439 and by the grant NS-6885.2010.2.


\end{document}